\documentclass[sigconf]{acmart}

\setcopyright{none}
\renewcommand\footnotetextcopyrightpermission[1]{}
\usepackage{array} 
\usepackage{multirow} 
\usepackage{csquotes}

\AtBeginDocument{%
  }

\setcopyright{acmlicensed}
\copyrightyear{2025}
\acmYear{2025}
\acmDOI{XXXXXXX.XXXXXXX}
\acmConference[CHI '25]{Late Breaking Work of the CHI Conference on Human Factors in Computing Systems}{April 26--May 1,
  2025}{Yokohama, Japan}
\acmISBN{978-1-4503-XXXX-X/2018/06}

\begin{document}

\title{``It Warned Me Just at the Right Moment": Exploring LLM-based Real-time Detection of Phone Scams}


\author{Zitong Shen}
\affiliation{%
  \institution{The Hong Kong Polytechnic University}
  \city{Hong Kong}
  \country{China}
}
\email{esther.shen@connect.polyu.hk}

\author{Sineng Yan}
\affiliation{%
  \institution{Shenzhen University}
  \city{Hong Kong}
  \country{China}
}
\email{yansineng2023@email.szu.edu.cn}

\author{Youqian Zhang}
\authornote{Corresponding author.}
\affiliation{%
  \institution{The Hong Kong Polytechnic University}
  \city{Hong Kong}
  \country{China}
}
\email{you-qian.zhang@polyu.edu.hk}

\author{Xiapu Luo}
\affiliation{%
  \institution{The Hong Kong Polytechnic University}
  \city{Hong Kong}
  \country{China}
}
\email{csxluo@comp.polyu.edu.hk}

\author{Grace Ngai}
\affiliation{%
  \institution{The Hong Kong Polytechnic University}
  \city{Hong Kong}
  \country{China}
}
\email{csgngai@comp.polyu.edu.hk}

\author{Eugene Yujun Fu}
\affiliation{%
  \institution{The Education University of Hong Kong}
  \city{Hong Kong}
  \country{China}
}
\email{eugenefu@eduhk.hk}

\begin{abstract}
Despite living in the era of the internet, phone-based scams remain one of the most prevalent forms of scams. 
These scams aim to exploit victims for financial gain, causing both monetary losses and psychological distress. 
While governments, industries, and academia have actively introduced various countermeasures, scammers also continue to evolve their tactics, making phone scams a persistent threat.
To combat these increasingly sophisticated scams, detection technologies must also advance. 
In this work, we propose a framework for modeling scam calls and introduce an LLM-based real-time detection approach, which assesses fraudulent intent in conversations, further providing immediate warnings to users to mitigate harm.
Through experiments, we evaluate the method's performance and analyze key factors influencing its effectiveness. 
This analysis enables us to refine the method to improve precision while exploring the trade-off between recall and timeliness, paving the way for future directions in this critical area of research.
\end{abstract}



\maketitle

\section{INTRODUCTION}
\label{sec:introduction}

In recent years, numerous news reports have highlighted the devastating impact of fraudulent phone calls, which have caused victims to suffer catastrophic financial losses. 
For instance, an 82-year-old retiree lost \$690,000 of their life savings to a phone scam~\cite{new2024deepfakes}. 
Unfortunately, such tragedies continue to occur at an alarming rate. 
The 2024 Global State of Scams Report~\cite{rogers2024international} reveals that phone scams siphoned off \$1.03 trillion globally over the past year, highlighting the severity of this issue.
The harm caused by phone scams extends beyond financial loss. 
Studies have shown the emotional and psychological toll on victims, including prolonged stress, anxiety, depression, and, in extreme cases, even suicide~\cite{hu2022btg, hu2024gat}. 

\begin{figure*}[t]
    \centering
    \includegraphics[width=1\textwidth]{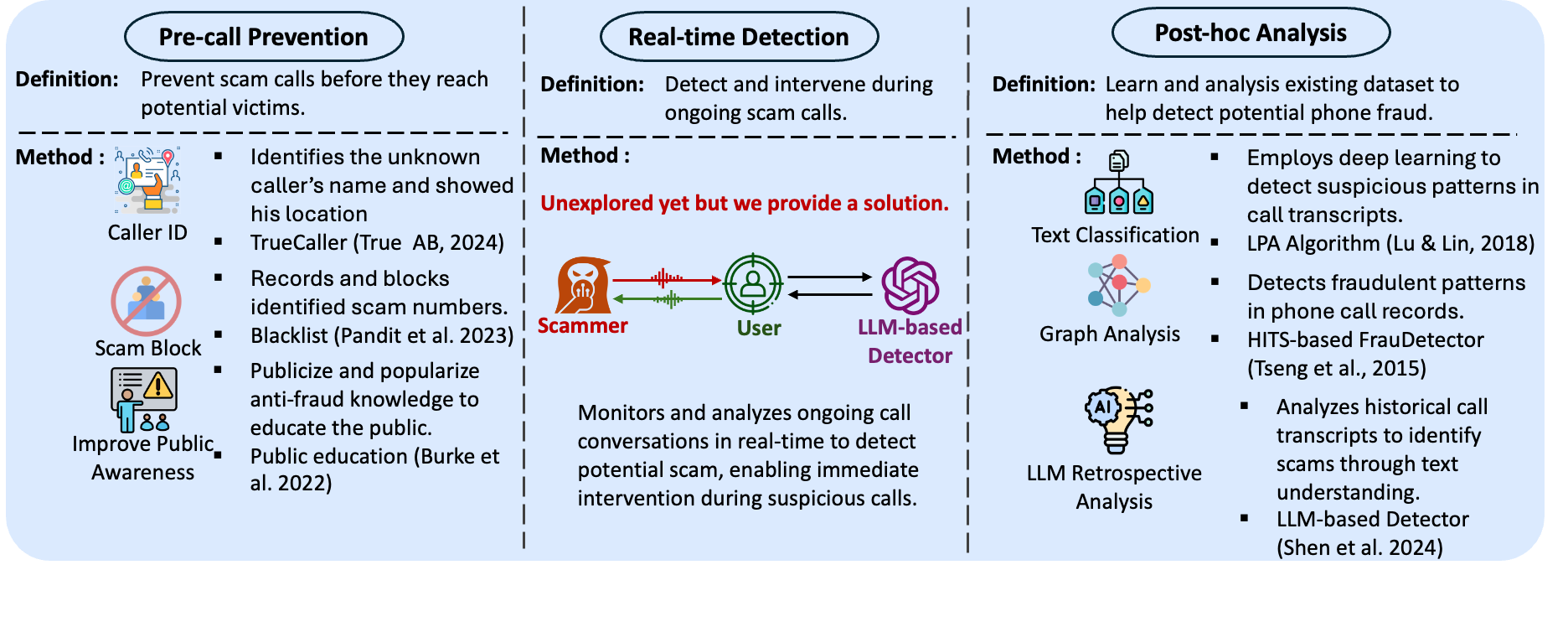}
    \caption{Taxonomy of Phone Scam Prevention Methods: A Temporal Perspective.}
    \label{figure1}
\end{figure*}

Governments, industries, and academia have implemented various measures to combat of phone scams, as summarized in the taxonomy of phone scam prevention methods in Fig.~\ref{figure1}.
Governments primarily focus on prosecuting criminal organizations and educating the public about fraud prevention~\cite{burke2022educational, deliema2020financial}. 
The industry has developed pre-call prevention methods, such as call-blocking applications~\cite{truecaller2024} and blacklists~\cite{pandit2023combating}, which attempt to filter out potential scam calls before they reach users. 
Meanwhile, academic research has explored post-call analysis techniques, employing AI tools to retrospectively identify scam patterns in call data~\cite{peng2018fraud,tseng2015fraudetector, chadalavada2024distinguishingscamsfraudensemble, chang2024exposingllmvulnerabilitiesadversarial, shen2024combatingphonescamsllmbased}.
Despite these efforts, the effectiveness of these approaches is increasingly undermined by the adaptability of scammers. 
Pre-call prevention methods, such as caller ID systems and blacklists, are vulnerable to caller ID spoofing and the use of VoIP services, which allow scammers to change their phone numbers frequently and evade detection~\cite{mustafa2018end, xing2020automated}. 
Likewise, public education campaigns, while valuable, cannot fully prevent individuals from falling victim to scams, as scammers exploit psychological manipulation to bypass even well-educated users' defenses~\cite{parti2023if}.
Post-call analysis, using methods such as text classification, graph analysis, and even LLM-based approaches,while useful for identifying patterns in scam calls after they occur, fails to provide timely intervention. 
In many cases, victims may transfer money or share sensitive information during the call itself~\cite{razaq2021we}.
These challenges highlight a critical gap in existing anti-scam strategies: the lack of effective real-time detection systems that can intervene during ongoing scam calls to prevent harm before it occurs.

To address this pressing gap, we propose a novel LLM-based real-time phone scam detection framework. 
Unlike pre-call and post-call approaches, our system analyzes phone conversations in real time during the call, enabling immediate detection of potential scams and providing timely warnings to users. 
This real-time capability is especially critical in situations where scammers apply psychological pressure, as early intervention can disrupt the scam process and prevent victims from making irreversible decisions.
Our approach builds on recent advancements in large language models (LLMs), which have demonstrated exceptional capabilities in understanding and analyzing human language. 
By leveraging LLMs, our framework can detect subtle linguistic cues and conversational anomalies indicative of scam behavior. 
Specifically, our work aims to address the following research questions:
\begin{itemize}
    \item \textbf{RQ1: How does our LLM-based real-time detection system perform in identifying phone scams?} To answer this question, we evaluate the performance of our framework using publicly available datasets and investigate its ability to detect scams during live phone conversations.

    \item \textbf{RQ2: What factors influence the method’s detection performance?} Through comprehensive analysis, we identify key factors that impact the system's effectiveness. These insights guide us in improving the framework to achieve a balance between precision and recall.
    
\end{itemize}
By answering these research questions, we aim to provide a comprehensive understanding of the capabilities and limitations of LLM-based real-time phone scam detection, optimize its performance for practical deployment, and evaluate its potential to safeguard users while delivering a seamless and effective intervention experience.

\section{SYSTEM MODELING AND DETECTION METHOD}
\label{sec:system_modeling_and_detection_method}

\begin{figure*}[t]
    \centering
    \includegraphics[width=\textwidth]{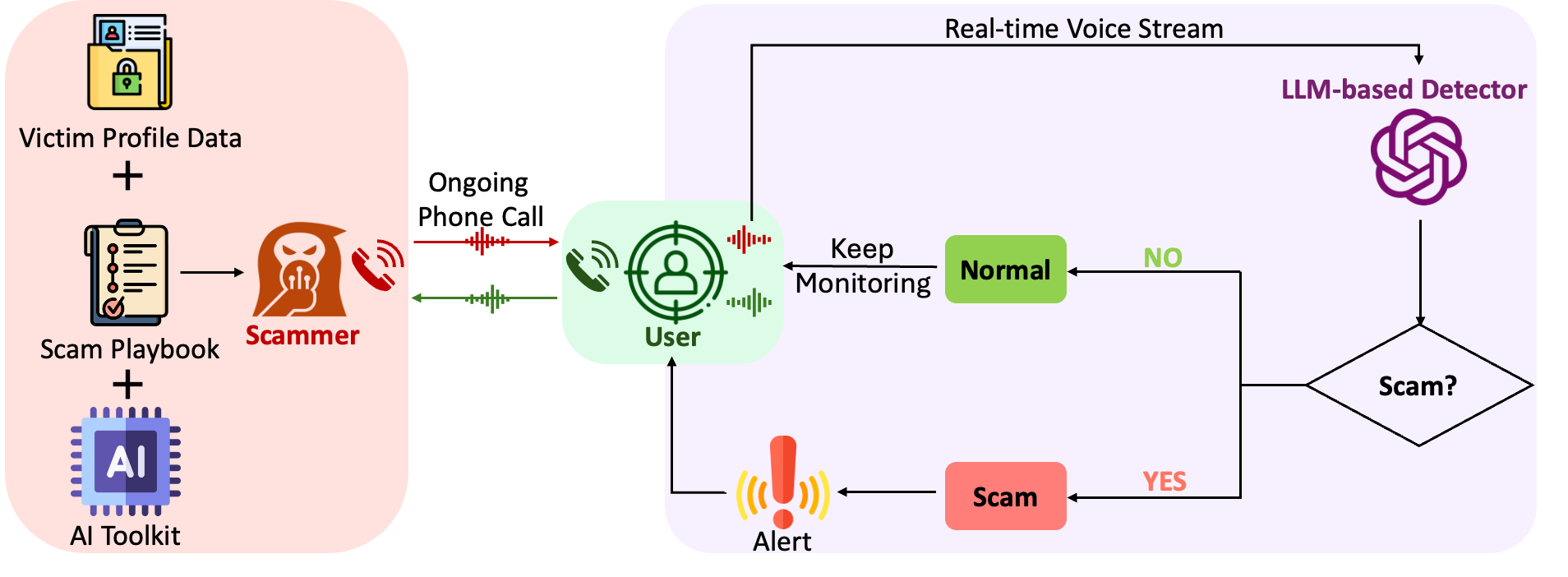}
    \caption{The system consists of three components, including ``Scammer'', ``User'', and ``LLM-based Detector''.}
    \label{fig:system_architecture}
\end{figure*}

In this section, we propose a high-level system model for phone scam detection, abstracted from reported cases/news. 
The model integrates our LLM-based detection method, which operates in real time to protect users during phone conversations. 

\subsection{System Model}

Below, we detail the roles, objectives, and capabilities of the three primary components of the system: the Scammer, the User, and the LLM-based Detector, as illustrated in Fig.~\ref{fig:system_architecture}.

\subsubsection{Scammer}
The scammer is the initiator of fraudulent calls, driven by financial gain and the exploitation of their targets. To achieve their goals, scammers employ a range of strategies, leveraging modern technology and psychological manipulation.
The scammer collects personal data about a target through public social media profiles, data breaches, or other sources. 
This information helps scammers build credibility and gain the trust needed to manipulate their victims effectively~\cite{gressel2024discussion}.
Relying on pre-defined scripts or playbooks, scammers can extract sensitive information, financial assets, or other valuable property.
Further, the scammer can enhance their deception with AI-powered technologies such as voice cloning tools which could impersonate loved ones, trusted authorities, or even political figures to manipulate victims~\cite{kennedy2014voice, cheyenne2024scammers}.
The scammer’s attack vector is through phone calls, exploiting the direct voice communication to manipulate victims.
Please note that in this work, we focus exclusively on audio-based phone scams and do not consider video call scams (e.g., deepfake-based attacks) as part of the system.

\subsubsection{User}
The User is the target of the scammer and is the primary beneficiary of the scam detection method.
The user’s primary objective is to avoid falling victim to scams while maintaining the ability to engage in normal phone communication. however, users do not assume every call is fraudulent, as such an assumption would significantly disrupt normal daily communication.
Nonetheless, several factors make users vulnerable to scams.
While users may have a basic awareness of phone scams and some level of risk aversion, they often lack the expertise required to identify sophisticated or evolving tactics. 
Emotional manipulation, including inducing panic or trust, is a key factor that makes users susceptible~\cite{mohamad2023anatomy}
Under the influence of the scammer, users may be coerced into disclosing sensitive information, transferring funds, or making decisions that conflict with their best interests.

\subsubsection{LLM-based Detector}
The LLM-based detector, operating on the user's phone, serves as the protective mechanism within the system. Its primary objective is to analyze ongoing phone conversations in real time, detect potential scams, and provide immediate intervention.
The detector continuously analyzes the content of the phone call through the text converted from the live voice stream, leveraging advanced LLM capabilities to identify suspicious patterns or conversational anomalies indicative of fraudulent activity. 
If no scam tendencies are detected, the detector remains unobtrusive, passively monitoring the conversation. However, when a scam pattern is identified, it generates an alert message to notify the user. 
This alert is conveyed through an appropriate medium (e.g., visual or audio notification) to warn the user of the potential danger. 
The timely alert acts as a critical intervention, allowing the user to terminate the call to avoid further potential harm.

\subsection{Detection Method}

The LLM-based detector performs fraud assessments by analyzing the conversation after each speaker's turn. 
When a participant, either the scammer or the user, completes their utterance, the detector evaluates the newly received input while maintaining contextual awareness of the prior conversational history.
This turn-by-turn evaluation ensures the timely detection of suspicious elements, leveraging the entire conversation history to make informed decisions. 

We utilize two distinct types of prompts in our LLM-based detector: a binary classification prompt and an enhanced prompt that incorporates an uncertainty category.
Real-time detection utilizing the binary classification prompt is referred to as ``RT'', while real-time detection leveraging the enhanced prompt with the uncertainty option is denoted as ``UNC''.

\subsubsection{Binary Classification Prompt (RT)}
Initially, the detector employs a straightforward binary classification approach to categorize the ongoing conversation as either ``FRAUD'' or ``SAFE''. The prompt used for this classification is as follows:

\begin{displayquote}
\texttt{Please analyze the call content and detect whether it is a fraud call. Please carefully analyze the suspicious features in the conversation. If it is a fraud call, please only return "FRAUD". If it is a normal call, please only return "SAFE". Do not return anything else.}
\end{displayquote}

\subsubsection{Prompt with Uncertain Option (UNC)}
\label{sec:prompt_with_uncertain_option}
While the binary classification framework is simple and intuitive, our experiments (discussed in later sections) reveal that it can lead to reduced performance. 
Specifically, the model may prematurely make decisions based on limited information, resulting in lower accuracy. 
To address this limitation, we introduce a third category, ``UNCERTAIN'', which allows the model to defer its decision when there is insufficient evidence to classify the call confidently. The updated prompt is as follows:

\begin{displayquote}
\texttt{Please analyze the call content and detect whether it is a fraud call. Please carefully analyze the suspicious features in the conversation. If it is a fraud call, please only return "FRAUD". If it is a normal call, please only return "SAFE". If there is insufficient information (e.g., it is not yet obvious that the fraud is present), please return "UNCERTAIN". Do not return anything else.}
\end{displayquote}

\section{DATASETS, LLM MODELS, and METRICS}

\subsection{Datasets}
To evaluate the effectiveness of our real-time detection approach, we employed open-source datasets that encompass both authentic and synthetic data sources~\cite{shen2024combatingphonescamsllmbased}. 
These datasets were carefully curated to include examples of both positive samples (scam phone calls) and negative samples (benign phone calls).
It is important to note that the datasets used in this study are in Chinese. For clarity and inclusivity, all transcript examples presented in the following sections have been translated into English. Additionally, any sensitive information within the transcripts has been anonymized to safeguard privacy.

\subsubsection{Authentic Dataset (Real)}
The authentic dataset comprises real-world transcripts of both scam and normal phone calls. 
These recordings were sourced from publicly available video platforms, such as YouTube. The audio data was processed using speech-to-text technology to generate the corresponding transcripts. 
This dataset reflects realistic conversational patterns, including diverse linguistic styles, emotional cues, and spontaneity, making it an essential component for evaluating real-world performance.

\subsubsection{Synthetic Dataset (Syn.)}
To complement the authentic dataset, a synthetic dataset was constructed. 
This dataset includes both scam and normal conversations generated to mimic real-world scenarios while intentionally embedding suspicious keywords or conversational elements. The synthetic dataset provides controlled examples enabling an assessment of the model’s ability to generalize across diverse and artificially constructed cases.

\subsection{LLM Models}
For our evaluation, we selected a diverse set of LLMs to implement the detector: GPT-4~\cite{achiam2023gpt4}, GPT-4o~\cite{hurst2024gpt4o}, GLM4~\cite{glm2024chatglm}, Doubao-Pro-32k~\cite{doubao-pro-32k}, and ERNIE-3.5-8k~\cite{ernie-3.5-8k}.
We selected these models to capture a broad spectrum of strengths, particularly in natural language understanding.
The diversity of these models allows us to evaluate the generalize ability of our detection method across different architectures, and benchmark state-of-the-art models to identify the most suitable candidates for real-world deployment.

\subsection{Metrics}
To comprehensively evaluate the performance of our detection system, we used the following metrics: accuracy, precision, recall, and F1-score.
Accuracy (Acc.) measures the proportion of correctly classified samples (both positive and negative) out of the total number of samples: $\text{Acc.} = \frac{\text{True Positives} + \text{True Negatives}}{\text{Total Samples}}$. Accuracy provides an overall measure of performance, the higher, the better.
Precision quantifies the proportion of correctly predicted positive samples (scam calls) out of all samples classified as positive. It is defined as: $\text{Prec.} = \frac{\text{True Positives}}{\text{True Positives} + \text{False Positives}}$.
High precision indicates the model’s ability to avoid false alarms, ensuring that flagged calls are genuinely scams.
Recall measures the proportion of actual positive samples (scam calls) that are correctly identified by the model. It is defined as: $\text{Rec.} = \frac{\text{True Positives}}{\text{True Positives} + \text{False Negatives}}$.
High recall ensures that most scam calls are detected, minimizing the risk of missed threats.
The F1-score (F1) is the harmonic mean of precision and recall, providing a balanced measure of the model’s ability to detect scams while minimizing false positives and false negatives.

\section{EXPERIMENTS \& ANALYSES}

In this section, we experimentally evaluate the performance of the proposed detection system and try to answer the research questions raised in Section~\ref{sec:introduction}.

\subsection{RQ1: How does our LLM-based real-time detection system perform in identifying phone scams?}

\begin{table}[t]
    \centering
    \setlength{\tabcolsep}{0.8pt}
    \footnotesize
    \begin{tabular}{ll|ccc|ccc|ccc|ccc|ccc}
        \toprule
        & & \multicolumn{3}{c|}{\textbf{GPT4}} & \multicolumn{3}{c|}{\textbf{GPT4o}} & \multicolumn{3}{c|}{\textbf{GLM4}} & \multicolumn{3}{c|}{\textbf{Doubao}} & \multicolumn{3}{c}{\textbf{ERNIE}} \\
        & & RT & UNC & RET & RT & UNC & RET & RT & UNC & RET & RT & UNC & RET & RT & UNC & RET \\
        \midrule
        \multirow{2}{*}{\textbf{Acc.}} & Real & 0.99 & 0.97 & 0.94 & 1.00 & 0.99 & 0.98 & 0.98 & 0.98 & 0.96 & 0.99 & 0.95 & 0.86 & 0.97 & 0.96 & 0.93 \\
        & Syn. & 0.85 & 0.87 & 0.99 & 0.84 & 0.84 & 0.98 & 0.79 & 0.89 & 0.91 & 0.96 & 0.95 & 0.97 & 0.79 & 0.89 & 0.98 \\
        \midrule
        \multirow{2}{*}{\textbf{F1}} & Real & 0.99 & 0.97 & 0.94 & 1.00 & 0.99 & 0.98 & 0.98 & 0.98 & 0.96 & 0.99 & 0.95 & 0.84 & 0.96 & 0.96 & 0.93 \\
        & Syn. & 0.87 & 0.88 & 0.99 & 0.86 & 0.86 & 0.98 & 0.83 & 0.90 & 0.91 & 0.97 & 0.95 & 0.97 & 0.83 & 0.90 & 0.98 \\
        \midrule
        \multirow{2}{*}{\textbf{Prec.}} & Real & 1.00 & 1.00 & 1.00 & 1.00 & 1.00 & 1.00 & 0.96 & 1.00 & 0.98 & 1.00 & 1.00 & 1.00 & 0.94 & 1.00 & 1.00 \\
        & Syn. & 0.77 & 0.80 & 1.00 & 0.76 & 0.77 & 0.98 & 0.70 & 0.83 & 0.95 & 0.94 & 1.00 & 1.00 & 0.70 & 0.85 & 1.00 \\
        \midrule
        \multirow{2}{*}{\textbf{Rec.}} & Real & 0.98 & 0.94 & 0.89 & 1.00 & 0.98 & 0.96 & 1.00 & 0.96 & 0.93 & 0.98 & 0.90 & 0.72 & 1.00 & 0.92 & 0.87 \\
        & Syn. & 1.00 & 0.98 & 0.98 & 1.00 & 0.98 & 0.98 & 1.00 & 0.98 & 0.86 & 0.98 & 0.90 & 0.94 & 1.00 & 0.94 & 0.97 \\
        \bottomrule
    \end{tabular}
    \caption{Performance comparison of different LLMs on real and synthetic datasets using three detection methods - RT: Real-Time Detection, UNC: Real-time Detection with Uncertain Option, RET: Retrospective Detection}
    \vspace{-20pt}
    \label{tab:detection_results}
\end{table}

We evaluate our proposed detection method using five LLMs previously mentioned on both authentic and synthetic datasets. 
For real-time detection, we initially utilize a binary classification prompt (i.e., RT). 
As a baseline for comparison, we include the results from previous work~\cite{shen2024combatingphonescamsllmbased}, where detection was performed using retrospective analysis, denoted as ``RET''. 
The evaluation results are presented in Table~\ref{tab:detection_results}.

In terms of recall, all five models exhibit exceptionally high performance, achieving recall rates ranging from 0.98 to 1.00 on both real and synthetic datasets. 
Notably, GPT4o, GLM4, and ERNIE achieve perfect recall scores of 1.00, demonstrating the system's high effectiveness in identifying the majority of potential scam calls in real-time scenarios, highlighting the detection method in capturing fraudulent activity without missing critical cases.

However, when considering precision, a significant disparity emerges between real-time detection and retrospective analysis. 
While retrospective analysis achieves precision rates ranging from 0.95 to 1.00, real-time detection exhibits noticeably lower precision values (0.70–0.77) across most models on the synthetic dataset. 
This gap suggests that, although real-time detection excels in identifying scam calls (high recall), it struggles with false positives, leading to reduced precision.

These findings motivate us to further explore RQ2 to identify and analyze the factors influencing the precision performance of our system during real-time detection. 
Addressing these factors is critical to enhance the overall reliability and practicality of real-time detection in real-world applications.

\begin{figure}[t]
    \centering
    \includegraphics[width=\columnwidth]{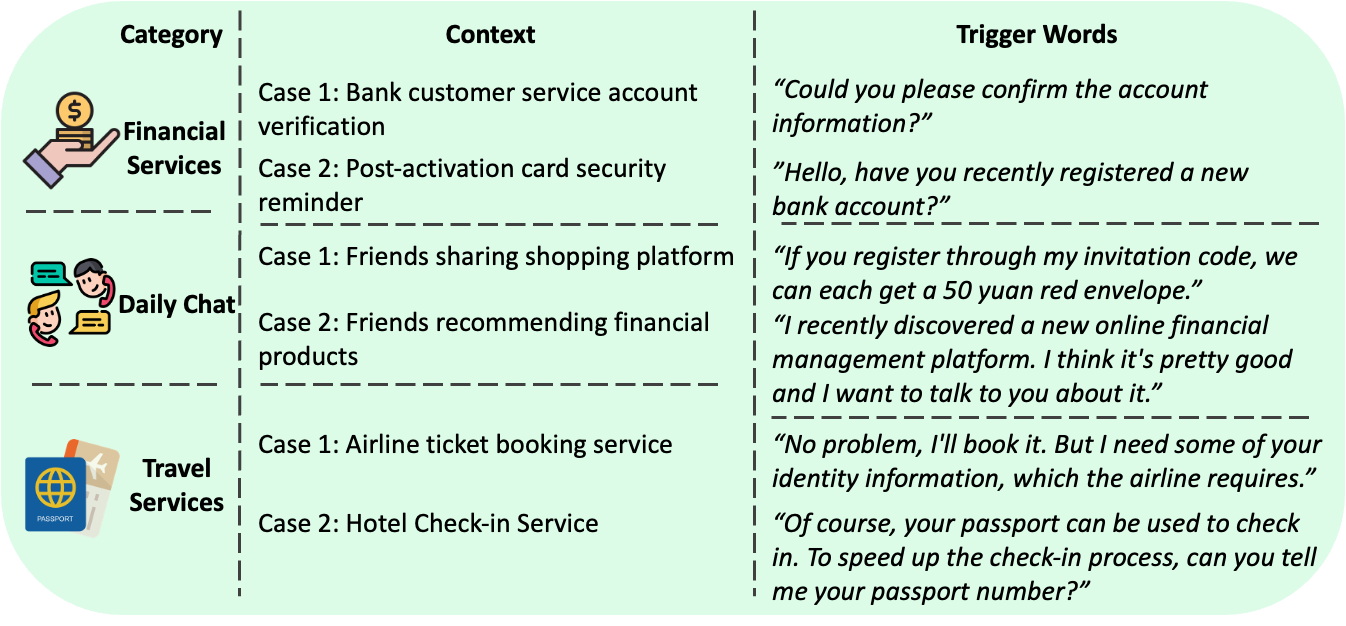}
    \caption{Analysis of false positive scenarios in fraud detection systems}
    \label{fig:false_alerts}
    \vspace{-5pt}
\end{figure}

\begin{figure*}[t]
    \centering
    \includegraphics[width=\textwidth]{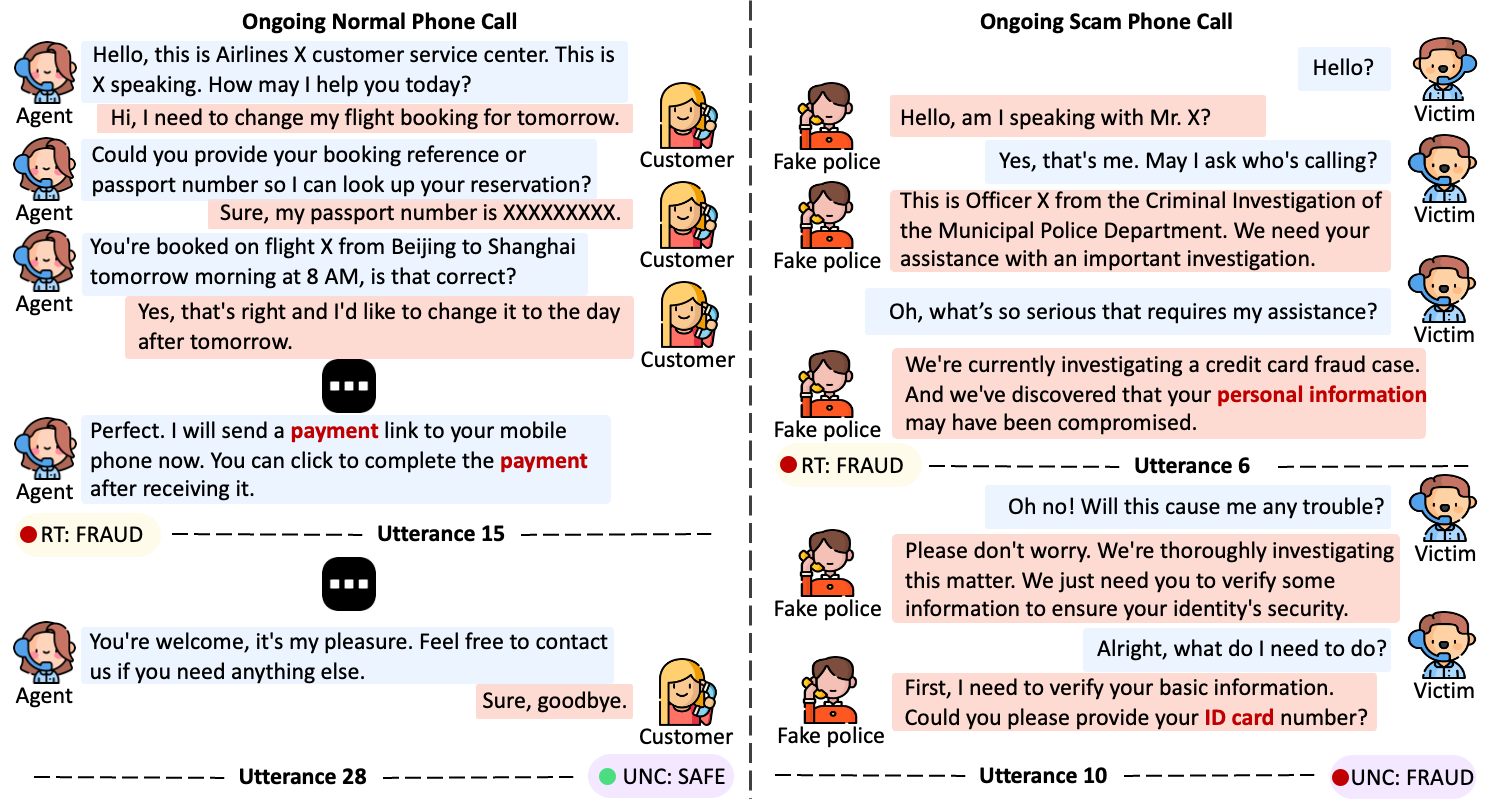}
    \caption{Visualization of fraud detection scenarios in an ongoing normal phone call and a scam phone call. Suspicious patterns are highlighted.}
    \label{fig:fraud_detection}
    \vspace{-5pt}
\end{figure*}

\subsection{RQ2: What factors influence the system’s detection performance?}

In real-time detection, the model processes each utterance incrementally during a phone conversation, 
with access only to the current utterance and a limited preceding context, as defined by the prompt design.
In contrast, retrospective analysis allows the model to examine the entire conversation history before making a judgment. 
This comprehensive view provides the model with complete contextual information, enabling it to interpret each utterance in light of the full dialogue, further improving its precision by reducing false positives.
Therefore, the difference in context availability between real-time detection and retrospective analysis may be one of the reasons that affect the rate of false positives.

By investigating the results, we identified a recurring issue: the detection method classifies the conversion as fraud if it contains certain suspicious keywords, e.g., payment, ID, classified as fraud-related terms in previous work~\cite{shen2024combatingphonescamsllmbased}.
Through our analysis of the system's false positive cases, as illustrated in Figure~\ref{fig:false_alerts}, we identified three main categories of legitimate scenarios that frequently trigger fraud alerts. In our observation, these interactions were frequently flagged as potential fraud due to their similarity with common scam patterns -- they all involve identity verification, financial information sharing, or urgent decision-making. Such false alert not only annoy users, but also could lead to users disabling or ignoring the protection system because they intend to mitigate undesired interruptions by suppressing alerts~\cite{li2023alert}. 

For instance, Figure~\ref{fig:fraud_detection} illustrates a normal customer service interaction where an agent assists a customer with re-booking a flight. 
The real-time detection method incorrectly flags the conversation as fraudulent at utterance 15, triggered by the agent’s use of the word ``payment'' while discussing the modification fee. 
Without access to the subsequent dialogue that clarifies the context, the model interprets the term as suggestive of fraud.

This phenomenon is particularly evident when comparing detection performance between the two datasets. 
While the models achieve high precision on the authentic dataset, where conversations naturally include contextual nuances, their precision drops significantly on the synthetic dataset, where normal dialogues are deliberately generated around suspicious keywords. This performance gap highlights the importance of contextual completeness in reducing false positives.

\subsection{Effectiveness of Involving Uncertain Option}

The observations discussed above motivate the introduction of a mechanism that allows the detector to delay making a definitive decision when uncertainty arises. 
This approach enables the system to gather additional conversational context before classifying. 
As introduced in Section~\ref{sec:prompt_with_uncertain_option}, the improved prompts with the uncertain option (denoted as ``UNC'') were specifically designed to address this challenge. 
The results of incorporating this option in real-time detection are summarized in Table~\ref{tab:detection_results}.

\subsubsection{Improved Precision}
The experimental results demonstrate that the use of the ``UNC'' leads to a significant improvement in precision, particularly on the synthetic dataset. For instance, the precision of Doubao increased from 0.70 to 0.83, and ERNIE rose from 0.70 to 0.85.
These improvements highlight the effectiveness of the ``UNC'' in reducing false positives, which may be commonly caused by incomplete information or an over-sensitivity to suspicious keywords as discussed above.

Figure~\ref{fig:fraud_detection} provides a concrete example of how ``UNC'' improves detection performance. 
For an ongoing phone call, after incorporating the uncertain response mechanism, the system correctly classified the conversation as ``SAFE'' at Utterance 28, the end of the conversation. 
Without ``UNC'', the same conversation might have been misclassified earlier due to incomplete context or isolated suspicious terms.

\subsubsection{Trade-offs in Recall and Timeliness}
The incorporation of ``UNC'' improves precision but comes at the cost of reduced recall. 
As shown in Table~\ref{tab:detection_results}, recall across all models drops from values between 0.98 and 1.00 to a range of 0.90 to 0.98 for both real and synthetic datasets. 
This trade-off indicates that while models become more cautious in making definitive fraud judgments, they may miss genuine fraudulent cases by labeling them as uncertain.

Also, there is a trade-off between detection reliability and timeliness. 
For example, in the scam phone call shown in Figure~\ref{fig:fraud_detection}, the fraudster impersonates a police officer investigating credit card fraud. 
The real-time system flags the call as fraudulent at Utterance 6 upon detecting phrases like ``personal information'' and ``credit card fraud''. 
However, ``UNC'' delays the fraud judgment until Utterance 10, when the impersonator explicitly requests ID card information. While this approach reduces false positives, it risks delaying critical alerts.

This delay introduces challenges in maintaining user trust. 
Early interventions may feel like the system is ``jumping to conclusions,'' whereas delayed warnings can be perceived as ``too late to be useful.'' 
Research has shown that user trust is strongly influenced by system accuracy and the timing of alerts ~\cite{dzindolet2003role,oduor2008effects,yu2017user, papenmeier2022complicated}. 
In our context, uncertainty in intervention timing may further burden users, forcing them to balance waiting for system confirmation with making timely decisions in their conversations.

\section{CONCLUSION AND FUTURE WORK}

This paper explored the application of Large Language Models (LLMs) for real-time phone scam detection, demonstrating their potential while uncovering key challenges through extensive experiments and analyses. 
Our system consistently achieved good performance across multiple LLMs, effectively identifying potential scams effectively and protecting users.
Meanwhile, our findings also underscore the inherent complexity of real-time fraud detection and outline several key directions for future work: addressing security and privacy concerns, as users may worry about fraud detection systems monitoring or storing personal data, and developing effective user notification strategies, which, though not covered in this paper, hold significant potential for improving user experience and trust.

While LLM-based real-time detection shows great promise, significant challenges remain in creating systems that effectively safeguard users without disrupting legitimate communications or eroding trust. 
Future research must address these challenges and will address a user study to explore the impact of the system on user experience, further refining user engagement strategies to maximize the real-world potential of LLMs.




\bibliographystyle{ACM-Reference-Format}
\bibliography{CHI}
\end{document}